\documentclass[10pt,aps,prl,twocolumn, longbibliography]{revtex4-2}

\usepackage{graphicx} 
\usepackage{amsmath}
\usepackage{amsfonts}
\usepackage{amssymb}
\usepackage[dvipsnames]{xcolor}

\usepackage{silence}
\usepackage[colorlinks=true]{hyperref}
\usepackage{graphicx}
\usepackage{tikz}
\usepackage{subfigure}
\usepackage{bbold}
\usepackage{braket}
\usepackage{microtype}
\usepackage{anyfontsize}
\usepackage{mathtools}

\hypersetup{
    linkcolor = BrickRed, 
    citecolor= ForestGreen, 
    urlcolor= RoyalBlue,
    pdftitle={Robuster Shattering}
}

\definecolor{myred}{HTML}{ed3624}
\definecolor{mygreen}{HTML}{13b205}
\definecolor{myblue}{HTML}{0379ee}
\definecolor{mypink}{HTML}{ff6ed8}
\definecolor{myyellow}{HTML}{fbe601}
\tikzset{
very thick/.style={line width=.7mm, line cap=round}
}

\DeclarePairedDelimiter\abs{\lvert}{\rvert}%
\DeclarePairedDelimiter\norm{\lVert}{\rVert}%

\makeatletter
\let\oldabs\abs
\def\abs{\@ifstar{\oldabs}{\oldabs*}}
\let\oldnorm\norm
\def\norm{\@ifstar{\oldnorm}{\oldnorm*}}
\makeatother

\usetikzlibrary{math}
\tikzmath{\c= cos(30); \s = sin(30);} 
\tikzset{mydash/.style={dash pattern=on .1cm off .1cm}}

\newcommand{\pardag}{\partial^\dagger\!}

\newcommand{\ZZ}{\mathbb{Z}}

\newcommand{\RK}{\mathrm{RK}}

\begin{document}

\title{Towards absolutely stable ergodicity breaking in two and three dimensions}

\author{Charles Stahl}
\altaffiliation{Present address: Department of Physics, Stanford University, Stanford, CA 94305}
\affiliation{Department of Physics and Center for Theory of Quantum Matter, University of Colorado Boulder, Boulder, Colorado 80309 USA}

\author{Oliver Hart}
\affiliation{Department of Physics and Center for Theory of Quantum Matter, University of Colorado Boulder, Boulder, Colorado 80309 USA}

\author{Rahul Nandkishore}
\affiliation{Department of Physics and Center for Theory of Quantum Matter, University of Colorado Boulder, Boulder, Colorado 80309 USA}
\date{October 10, 2024}

\begin{abstract}
We propose new physically reasonable 
systems capable of avoiding ergodicity at infinite time in the thermodynamic limit, even with generic perturbations and when coupled to a heat bath. 
In two dimensions, the \emph{rainbow loop soup} has (stretched) exponentially numerous
absolutely stable nonergodic states with diverging energy but vanishing energy density. 
In three dimensions the \emph{rainbow membrane soup} has (stretched) exponentially numerous nonergodic states with diverging energy barriers, leading to infinite-time robust ergodicity breaking that even survives coupling to a nonzero temperature heat bath. 
We describe our results in the language of exact emergent symmetries and demonstrate how the systems avoid common instabilities. Our construction naturally connects to quantum dimer models, topologically ordered systems, the group word construction, and Hamiltonians whose low-energy eigenstates exhibit anomalous entanglement entropy.
\end{abstract}

\maketitle

The ergodic hypothesis in quantum statistical mechanics postulates that over time many-body quantum dynamics explores all symmetry-allowed regions of Hilbert space. Here, `symmetry' includes higher-form symmetries~\cite{Nussinov2007, Gaiotto2015, mcGreevy2023generalized}, so topologically ordered systems are indeed ergodic after resolving the symmetry sectors.
Systems that violate the ergodic hypothesis robustly are a great white whale of quantum statistical mechanics. While affirmative constructions are known in infinite dimensions
(see, e.g., Ref.~\cite{YNL}), we focus on systems in three or fewer spatial dimensions, where the situation is murkier. 

{\it Integrable} systems violate the ergodic hypothesis~\cite{Baxter}, yet lack any argument for robustness. {\it Many-body localized} systems~\cite{MBLARCMP, MBLRMP} are believed to robustly break ergodicity at nonzero energy density, but require perfect isolation (i.e., no heat bath~\cite{mblbath}), and the ergodicity breaking does not persist to infinite times, except perhaps in one dimension~\cite{Imbrie}. Can one do better? Two parallel strands of research hold promise. One involves {\it Hilbert space shattering} (a.k.a.~fragmentation)~\cite{khemani2020shattering, Sala2020}, wherein certain symmetries cause Hilbert space to split into disconnected Krylov sectors even within each symmetry sector, giving symmetry-resolved ergodicity breaking at nonzero energy density. In presently known models this shattering does not survive generic perturbations, although when the parent symmetry is higher form, it can persist to timescales exponentially long in inverse perturbation strength even in the presence of nonlocal asymmetric perturbations~\cite{stephen2022ergodicity, stahl2023topologically, Balasubramanian2024glassy, khudorozhkov2024robust}. In parallel, {\it fracton} systems (see, e.g., Refs.~\cite{fractonarcmp, fractonrmp} for reviews) 
exhibit provably robust ergodicity breaking to infinite time at zero energy density in three dimensions~\cite{kim2016localization}, although the timescale becomes finite at nonzero energy densities~\cite{prem2017glassy, siva2017marginally}.

In this Letter, we propose a route to infinite-time symmetry-resolved ergodicity breaking in two and three dimensions, which survives generic perturbations and even coupling to a heat bath. Our constructions combine the `topologically fragmented' models of~\cite{stephen2022ergodicity, stahl2023topologically, Balasubramanian2024glassy, khudorozhkov2024robust} with the recently developed notion of {\it exact emergence of higher-form symmetries} \cite{pace2023exact}. In two dimensions, we argue that our construction yields robust infinite-time ergodicity breaking at zero energy density, similar to Ref.~\cite{kim2016localization} but in one fewer dimension. 
In three dimensions, we argue that it provides the first finite-dimensional example of infinite-time symmetry-resolved ergodicity breaking that is stable to generic perturbations and coupling to a nonzero-temperature heat bath.

\begin{figure}
\centering
\begin{tikzpicture}[scale=.5]

\def\maxy{3}
\def\maxx{6}

\begin{scope}
\clip (-\c-.03, -1) rectangle (6*\c+.03, 3+3*\s);
\foreach \x in {0,2,...,\maxx}
    \foreach \y in {0,2,...,\maxy}
        \draw (\x*\c-\c, \y*\s+\s+\y) -- (\x*\c, \y*\s+\y) -- (\x*\c+\c, \y*\s+\s+\y) (\x*\c, \y*\s+\y) -- (\x*\c, \y*\s+\y-1);

\pgfmathsetmacro{\upperbound}{\maxy + 1}
\foreach \x in {-2,0,...,\maxx}
    \foreach \y in {1,3,...,\upperbound}
        \draw (\x*\c, \y*\s+\s+\y) -- (\x*\c+\c, \y*\s+\y) -- (\x*\c+2*\c, \y*\s+\s+\y) (\x*\c+\c, \y*\s+\y) -- (\x*\c+\c, \y*\s+\y-1);
\end{scope}

\foreach \x in {0,2,...,\maxx}
    \draw[mydash] (\x*\c-\c, -1-\s) -- (\x*\c, -1) 
                  (\x*\c, -1) -- (\x*\c+\c, -1-\s) 
                  (\x*\c-\c, 3+3*\s) -- (\x*\c, 3+4*\s)
                  (\x*\c, 3+4*\s)-- (\x*\c+\c, 3+3*\s) ;

\foreach \y in {0,2,...,\maxy} {
    \draw[mydash] ( -\c, \y*\s+\s+\y) -- (-2*\c, \y*\s+\y) 
                  (-2*\c, \y*\s+\y-1) -- (-\c, \y*\s+\y-1-\s);d
    \draw[mydash] (6*\c, \s*\y+\y) -- (7*\c, \y*\s+\y+\s) 
                  (7*\c, \y*\s+\y+\s+1) -- (6*\c, \y*\s+\y+1+2*\s);
}

\draw[mydash] (-2*\c, 3+4*\s) -- (-\c, 3+3*\s);

\draw[very thick, ForestGreen] (-\c, 2+3*\s) -- (0, 2+2*\s) -- (\c, 2+3*\s) -- (2*\c, 2+2*\s) -- (2*\c, 1+2*\s) -- (\c, 1+\s) -- (0, 1+2*\s) -- (-\c, 1+\s) (6*\c, 2+2*\s) -- (6*\c, 1+2*\s);
\draw[very thick, ForestGreen, mydash] (-\c, 2+3*\s) -- (-2*\c, 2+2*\s) (-2*\c, 1+2*\s) -- (-\c, 1+\s);
\draw[very thick, ForestGreen, mydash] (7*\c, 2+3*\s) -- (6*\c, 2+2*\s) (6*\c, 1+2*\s) -- (7*\c, 1+\s);

\draw[very thick, RoyalBlue] (6*\c, -1) -- (6*\c, 0) -- (5*\c, \s) -- (5*\c, 1+\s) -- (4*\c, 1+2*\s) -- (3*\c, 1+\s) -- (3*\c, \s) -- (2*\c, 0) -- (2*\c, -1);
\draw[very thick, RoyalBlue, mydash] (2*\c, -1) -- (3*\c, -1-\s) (3*\c, -1-\s) -- (4*\c, -1) (4*\c, -1) -- (5*\c, -\s-1) (5*\c, -\s-1) -- (6*\c, -1);
\draw[very thick, RoyalBlue, mydash] (2*\c, 3+4*\s) -- (3*\c, 3+3*\s) (3*\c, 3+3*\s) -- (4*\c, 3+4*\s) (4*\c, 3+4*\s) -- (5*\c, 3+3*\s) (5*\c, 3+3*\s) -- (6*\c, 3+4*\s);

\end{tikzpicture}\hfill
\begin{tikzpicture}[scale=.5]

\def\maxy{3}
\def\maxx{6}

\begin{scope}
\clip (-\c-.03, -1) rectangle (6*\c+.03, 3+3*\s);
\foreach \x in {0,2,...,\maxx}
    \foreach \y in {0,2,...,\maxy}
        \draw (\x*\c-\c, \y*\s+\s+\y) -- (\x*\c, \y*\s+\y) -- (\x*\c+\c, \y*\s+\s+\y) (\x*\c, \y*\s+\y) -- (\x*\c, \y*\s+\y-1);

\pgfmathsetmacro{\upperbound}{\maxy + 1}
\foreach \x in {-2,0,...,\maxx}
    \foreach \y in {1,3,...,\upperbound}
        \draw (\x*\c, \y*\s+\s+\y) -- (\x*\c+\c, \y*\s+\y) -- (\x*\c+2*\c, \y*\s+\s+\y) (\x*\c+\c, \y*\s+\y) -- (\x*\c+\c, \y*\s+\y-1);
\end{scope}

\foreach \x in {0,2,...,\maxx}
    \draw[mydash] (\x*\c-\c, -1-\s) -- (\x*\c, -1) 
                  (\x*\c, -1) -- (\x*\c+\c, -1-\s) 
                  (\x*\c-\c, 3+3*\s) -- (\x*\c, 3+4*\s)
                  (\x*\c, 3+4*\s)-- (\x*\c+\c, 3+3*\s) ;

\foreach \y in {0,2,...,\maxy} {
    \draw[mydash] ( -\c, \y*\s+\s+\y) -- (-2*\c, \y*\s+\y) 
                  (-2*\c, \y*\s+\y-1) -- (-\c, \y*\s+\y-1-\s);d
    \draw[mydash] (6*\c, \s*\y+\y) -- (7*\c, \y*\s+\y+\s) 
                  (7*\c, \y*\s+\y+\s+1) -- (6*\c, \y*\s+\y+1+2*\s);
}

\draw[mydash] (-2*\c, 3+4*\s) -- (-\c, 3+3*\s);

\def\thisx{-2}

\draw[very thick, BrickRed, mydash] (\thisx*\c, 3+4*\s) -- (\thisx*\c + \c, 3+3*\s) (\thisx*\c + \c, 2+3*\s) (\thisx*\c + \c, 2+3*\s) -- (\thisx*\c, 2+2*\s) (\thisx*\c, 1+2*\s) -- (\thisx*\c + \c, 1+\s) ;
\draw[very thick, BrickRed] (\thisx*\c + \c, 3+3*\s) -- (\thisx*\c + \c, 2+3*\s) (\thisx*\c + \c, 1+\s) -- (\thisx*\c + \c, \s);
\draw[very thick, BrickRed, mydash] (\thisx*\c + \c, \s) -- (\thisx*\c, 0) (\thisx*\c, -1) -- (\thisx*\c + \c, -1-\s);

\def\thisx{0}

\draw[very thick, RoyalBlue, mydash] (\thisx*\c, 3+4*\s) -- (\thisx*\c + \c, 3+3*\s);
\draw[very thick, RoyalBlue] (\thisx*\c + \c, 3+3*\s) -- (\thisx*\c + \c, 2+3*\s) -- (\thisx*\c, 2+2*\s) -- (\thisx*\c, 1+2*\s) -- (\thisx*\c + \c, 1+\s) -- (\thisx*\c + \c, \s) -- (\thisx*\c, 0) -- (\thisx*\c + \c, \s) -- (\thisx*\c, 0) -- (\thisx*\c, -1);
\draw[very thick, RoyalBlue, mydash] (\thisx*\c, -1) -- (\thisx*\c + \c, -1-\s);

\def\thisx{2}

\draw[very thick, BrickRed, mydash] (\thisx*\c, 3+4*\s) -- (\thisx*\c + \c, 3+3*\s);
\draw[very thick, BrickRed] (\thisx*\c + \c, 3+3*\s) -- (\thisx*\c + \c, 2+3*\s) -- (\thisx*\c, 2+2*\s) -- (\thisx*\c, 1+2*\s) -- (\thisx*\c + \c, 1+\s) -- (\thisx*\c + \c, \s) -- (\thisx*\c, 0) -- (\thisx*\c + \c, \s) -- (\thisx*\c, 0) -- (\thisx*\c, -1);
\draw[very thick, BrickRed, mydash] (\thisx*\c, -1) -- (\thisx*\c + \c, -1-\s);

\def\thisx{4}

\draw[very thick, RoyalBlue, mydash] (\thisx*\c, 3+4*\s) -- (\thisx*\c + \c, 3+3*\s);
\draw[very thick, RoyalBlue] (\thisx*\c + \c, 3+3*\s) -- (\thisx*\c + \c, 2+3*\s) -- (\thisx*\c, 2+2*\s) -- (\thisx*\c, 1+2*\s) -- (\thisx*\c + \c, 1+\s) -- (\thisx*\c + \c, \s) -- (\thisx*\c, 0) -- (\thisx*\c + \c, \s) -- (\thisx*\c, 0) -- (\thisx*\c, -1);
\draw[very thick, RoyalBlue, mydash] (\thisx*\c, -1) -- (\thisx*\c + \c, -1-\s);

\def\thisx{6}

\draw[very thick, BrickRed, mydash] (\thisx*\c, 3+4*\s) -- (\thisx*\c + \c, 3+3*\s) (\thisx*\c + \c, 2+3*\s) -- (\thisx*\c, 2+2*\s) (\thisx*\c, 1+2*\s) -- (\thisx*\c + \c, 1+\s) ;
\draw[very thick, BrickRed] (\thisx*\c, 2+2*\s) -- (\thisx*\c, 1+2*\s)  (\thisx*\c, 0) -- (\thisx*\c, -1);
\draw[very thick, BrickRed, mydash] (\thisx*\c + \c, \s) -- (\thisx*\c, 0) (\thisx*\c, -1) -- (\thisx*\c + \c, -1-\s);

\end{tikzpicture}
\caption{Two constraint-satisfying configurations. Dashed lines denote edges that appear on both sides of the periodic boundary conditions. The two configurations have the same symmetry numbers but different Krylov labels.}
\label{fig:symloops}
\end{figure}
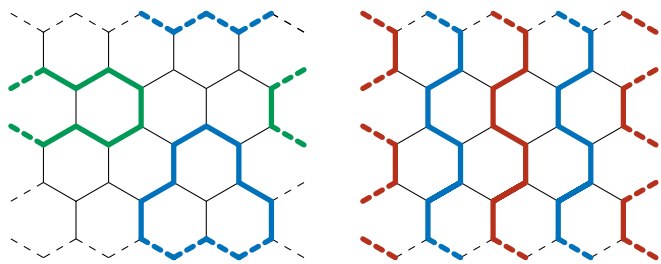

\emph{The model in two dimensions.}---%
We consider a honeycomb lattice with periodic boundary conditions where each horizontal slice cuts $L$ vertical bonds. On every edge, define a local Hilbert space spanned by states $\{|\alpha \rangle; \, \alpha = 0, \dots, m\}$. We will refer to $\alpha = 1,\dots, m$ as the nontrivial \emph{colors}, and $\alpha=0$ as the trivial or background color.
Define the operators
\begin{equation}
Z^{(\alpha)} = 1 - 2\ket{\alpha} \bra{\alpha}; \qquad 
X^{(\alpha)} = \ket{0}\bra{\alpha} + \ket{\alpha} \bra{0},
\end{equation}
which reduce to the ordinary Pauli operators when $m=1$ and $\alpha = 1$.
Furthermore, define the operators
\begin{equation}
A_v^{(\alpha)} = \prod_{e \in \pardag v} Z^{(\alpha)}_e,
\end{equation}
which act as $+1$ if vertex $v$ has an even number of incident edges $e$ in state $\alpha$, and $-1$ otherwise. 

The $A_v^{(\alpha)}$ operators commute, but 
cannot be satisfied (i.e., $A_v^{(\alpha)} = 1$) simultaneously for all $\alpha$ because every vertex has three incident edges. 
Instead, if we ignore the background color ($\alpha=0$), then the $A_v^{(\alpha)}$ can be simultaneously satisfied for $\alpha = 1,\dots,m$ and the classical Hamiltonian
\begin{equation}
H_\mathrm{cl} = \frac{J}{2} \sum_v \sum_{\alpha=1}^m \Big(1 - A_v^{(\alpha)} \Big), \label{eqn:constraint}
\end{equation}
penalizes constraint violations ($J > 0$). Hilbert space splits into a low-energy subspace that obeys all of the constraints and a high-energy subspace that does not. Graphically, in the low-energy states, the edges in state $\alpha=1,\dots, m$ form closed loops, while the edges in state $0$ form the vacuum, as illustrated in Fig.~\ref{fig:symloops}. In the language of Refs.~\cite{Balasubramanian2024glassy,khudorozhkov2024robust}, the above construction defines a $\ZZ_2^{*m}$ group loop model. 

Local constraint-preserving dynamics allow small loops to nucleate and fluctuate. For every face $f$ we introduce the kinetic operator 
\begin{align}
B_f^{K,\alpha} = \prod_{e \in \partial f} X_e^{(\alpha)},
\end{align}
which implements these local dynamics and therefore commutes with the constraint. Isolated noncontractible loops cannot be created via a local process without first creating an open loop and violating the constraint. Meanwhile, loops of different color cannot cross each other because only three edges meet at each vertex. Thus, if the constraints in Eq.~\eqref{eqn:constraint} are exact, every product state in the computational basis has an immutable \emph{label} $\lambda$ extracted from the pattern of noncontractible loops. To extract the label, take the ordered sequence of colors of the noncontractible loops, and delete repetitions (as in the 1d pair-flip model~\cite{caha2018pairflip, moudgalya2022hilbert}) to obtain a `word' with no repeated letters~\footnote{In order to obtain labels in one-to-one correspondence with Krylov sectors, identify labels that are equivalent up to periodic translations, unless the word has length $L$. See Ref.~\cite{khudorozhkov2024robust} for further details.}. There are $\sim (m-1)^L$ such words, which are constants of motion and which therefore label $\sim (m-1)^L$ dynamically disconnected Krylov sectors, which we call $\mathcal{K}_\lambda$. 
We define $\abs{\lambda}$ as the number of letters in the label $\lambda$, satisfying $0 \le |\lambda| \le L$. The challenge is to show that this `shattering' of Hilbert space persists when the constraint is implemented energetically, as in Eq.~\eqref{eqn:constraint}.

We now define the potential operator
\begin{equation}
B_f^{P,\alpha} = \prod_{e\in \partial f} \Big( \ket{0}\bra{0} + \ket{\alpha}\bra{\alpha} \Big)_e,
\end{equation}
which projects onto states on which $B_f^{K,\alpha}$ acts nontrivially. Within the space of constraint-satisfying states, we can define the Rokhsar-Kivelson (RK) \cite{RK} Hamiltonian
\begin{equation}
H_\RK[V,K] = \frac{1}{2}\sum_f \sum_{\alpha = 1}^m  \left( V B_f^{P,\alpha} - K B_f^{K,\alpha}\right). \label{eqn:HHF}
\end{equation}
Returning to the full tensor-product Hilbert space and tuning to the RK point $V=K$, we now consider the (unperturbed) Hamiltonian
\begin{align}
H_0 &= H_\mathrm{cl} + H_\RK[V=K, K] \label{eqn:H0} \nonumber\\
&= J \sum_v \sum_{\alpha = 1}^m \frac{1 \! - \! A_v^{(\alpha)}}{2} + K \sum_f \sum_{\alpha = 1}^m \frac{B_f^{P,\alpha} \! - \! B_f^{K,\alpha}}{2}. 
\end{align}
This Hamiltonian is written as a positive sum of projectors, so that any state annihilated by every projector has zero energy and is a ground state. 

In fact, for any Krylov sector $\mathcal{K}_\lambda$, the `RK' state corresponding to label $\lambda$, 
\begin{equation}
\ket{\Psi_\lambda} = \frac{1}{\sqrt{\left| \mathcal{K}_\lambda \right|}} \sum_{\psi \in \mathcal{K}_\lambda} \ket{\psi}, \label{eqn:groundstate}
\end{equation}
is annihilated by every projector in $H_0$~\eqref{eqn:H0} and is therefore a ground state~\cite{henley2004RK}.
The sum runs over all $Z$-basis states $\psi$ with label $\lambda$. If $\lambda$ is a maximal label, i.e., $|\lambda|=L$, then $\mathcal{K}_\lambda$ is a Krylov sector with only a single state and $\ket{\Psi}_\lambda$ is a product state. Alternatively, if $\lambda$ is short, 
then there are many states in $\mathcal{K}_\lambda$ and $\ket{\Psi_\lambda}$ is a macroscopic superposition of nonintersecting loops. We call a label short if $|\lambda| = o(L)$, and long otherwise. We emphasize that the macroscopic-superposition nature of the short-label states $|\Psi\rangle_{\lambda}$ will be crucial to endowing our model with greater stability than the models in Refs.~\cite{stephen2022ergodicity, stahl2023topologically, Balasubramanian2024glassy, khudorozhkov2024robust}.

Because of the nonintersecting colored loops,  we call this model the rainbow loop soup. 
This name also references the bicolor loop model of Ref.~\cite{zhang2024bicolor}, to which the rainbow loop soup reduces at $m=2$. The bicolor loop model has interesting entanglement properties, which the rainbow loop soup~\eqref{eqn:H0} should share. The rainbow loop soup also reduces to the 2d toric code~\cite{kitaev2003toric} at $m=1$.

We now deform away from the RK point $K=V$ by adding $\delta H = (V-K) \sum_f \sum_{\alpha=1}^m B_f^{P,\alpha} $ to Eq.~\eqref{eqn:H0}. This lifts the degeneracy between RK states $\ket{\Psi_\lambda}$ with different label lengths $|\lambda|$. As long as $\epsilon \equiv V-K$ is small, we can treat $\delta H$ as a perturbation to deduce properties of the true ground states. 
We focus on $\epsilon < 0$, when the true ground state is $\ket{\Psi_\varnothing}$, the macroscopic superposition of all states with the empty label $\varnothing$. However, other RK states with short labels locally look very similar to $\ket{\Psi_\varnothing}$ due to the large number of basis states in these Krylov sectors.
There are $(m-1)^{o(L)}$ short-label RK states and we will argue that all of these have 
subextensive energy relative to the true ground state.
In what follows, we call their perturbation-theoretic descendants \emph{steady states} because they have a definite label and remain invariant under the dynamics.

\emph{Exact emergent symmetry.}---%
The central claim of this letter is that, for a finite range of $\epsilon <0$ and for sufficiently small arbitrary (but $k$-local) perturbations, the perturbation-theoretic descendants of the short-label states remain steady states with diverging lifetimes in the thermodynamic limit. 
Our primary argument in favor of this claim comes from the recently established concept of exact emergent symmetry~\cite{pace2023exact}. The relevant result is that discrete 1-form symmetries in two dimensions are exact when they are emergent. Thus, our claim reduces to a claim that there is an emergent discrete 1-form symmetry (which is therefore exact), and the exact symmetry implies that the short-label states are steady states. We will argue that this follows because changing the Krylov sector label requires moving loop endpoints across the system~\cite{stahl2023topologically}, and loop endpoints violate the emergent higher-form symmetry. 

We now identify the key (exact emergent) symmetry. 
Along any dual path $\mathcal{P}'$ on the lattice, define the operator
\begin{equation}\label{eqn:exactsymm}
A_\mathcal{P'}^{(\alpha)} = \prod_{e\in \mathcal{P}'}Z_e^{(\alpha)},
\end{equation}
which evaluates to $+1$ if it crosses an even number of loops of color $\alpha$ and $-1$ otherwise. If $\mathcal{P}'$ is contractible, then $A_\mathcal{P'}^{(\alpha)}$ is a product of $A_v^{(\alpha)}$ operators and acts as $+1$ on any configuration allowed by the constraints. However, if $\mathcal{P}'$ is noncontractible, then $A_\mathcal{P'}^{(\alpha)}$ is a nontrivial $\ZZ_2$-valued operator. Taken together, the operators generate a $\ZZ_2^m$ 1-form symmetry, which distinguishes $2^{2m}$ symmetry sectors on a torus, labeled by the parity of the number of noncontractible loops of each color. Meanwhile, there are $\sim (m-1)^L$ Krylov sectors, so that some states from the same symmetry sector must belong to different Krylov sectors. For example, the two configurations in Fig.~\ref{fig:symloops} have the same symmetry numbers but different Krylov labels. When the symmetry is exact, the Krylov sectors are disconnected. There are operators that commute with the symmetry and change the Krylov label, 
but these operators have diverging support (since the label is topologically robust \cite{ stahl2023topologically}) and so cannot be generated by any $k$-local perturbation without going through $O(L)$ intermediate states that violate the symmetry (which is exact by postulate).

Once we consider perturbations, the symmetries cease to be exact microscopically. For example, the perturbation $X^{(\alpha)}_e$ fails to commute with any $A_\mathcal{P'}^{(\alpha)}$ where $e \in \mathcal{P}'$. However, within the low-energy subspace of $H_0$, the symmetry is emergent at every order in perturbation theory up to order $L$. This is because the only on-shell operators that fail to commute with the symmetry are the ones that insert noncontractible loops into the system. Thus, the arguments of Ref.~\cite{pace2023exact} tell us that the symmetry is exact in the IR, and that the short-label RK states remain steady states of the perturbed Hamiltonian. This does not imply that the RK states with longer labels are steady states, because in two dimensions exact emergence is limited to zero temperature \cite{pace2023exact}, whereas long-label states are already at nonzero energy density relative to the true ground state, which would correspond to nonzero temperature if they thermalized (and we cannot assume that they do not thermalize, since that would be assuming what we wish to prove).

\emph{Avoiding common instabilities.}---%
We now argue that the rainbow loop soup is immune to some common instabilities, starting with false-vacuum decay. Recall that, for $\epsilon < 0$, the unique true ground state is $\ket{\Psi_\varnothing}$ and the other RK states are false vacua~\footnote{Strictly the RK states are the steady states of the RK Hamiltonian $H_0$, not the perturbed Hamiltonian $H$. The true false vacua will be connected to these RK states by a Schrieffer-Wolff transformation \cite{SW}.}. Could there be false vacuum decay, where some operator is able to create a bubble of true vacuum that then expands to fill the whole system? In the 2d Ising model these vacuum bubbles have an energy benefit proportional to their area and energy cost proportional to their perimeter, so the onshell bubble has a finite size and false vacuum decay happens in finite time (as witnessed by local observables). A similar false-vacuum-decay instability plagues the models of, e.g., Refs.~\cite{stahl2023topologically,Balasubramanian2024glassy,khudorozhkov2024robust}. However, we now argue that the rainbow loop soup evades this instability. 

In the rainbow loop soup, states with different label length have different energy density $\rho$. 
In the Supplement~\cite{supplement} we provide numerical evidence that suggests that $\rho$ only depends on the combination $|\lambda| / L$, and further that $\rho(|\lambda| / L) \rightarrow 0$ as $|\lambda| / L \rightarrow 0$ \footnote{However, this is by no means proven, and placing this (highly plausible) result on firm footing is a key step towards a rigorization of our results}. This makes intuitive sense, since in the limit $|\lambda|/L \rightarrow 0$ the noncontractible loops comprising the label have zero density, and thus do not interact. It then follows that every short-label false vacuum has vanishing energy density in the thermodynamic limit. 
Intuitively, vanishing energy density should lead to a diverging lifetime, due to the diverging size of the minimal `onshell bubble' required to connect vacua. 
This intuition has recently been shown to be true rigorously in the simpler setting of Ising models~\cite{YSL}, and we anticipate that a similar rigorization may be possible in the present setting. 

In the spin-liquid setting, the other potential instability to worry about is the instanton proliferation that plagues 2d U(1) spin liquids~\cite{Polyakov}. In fact, the rainbow loop soup looks like the 2d U(1) spin liquid in that both have a diverging number of ground states at their RK points. However, the instanton instability in U(1) spin liquids is a fluctuation in the emergent U(1) gauge field~\cite{Polyakov}, which is related to that model's U(1) higher-form symmetry. The rainbow loop soup has a discrete higher-form symmetry instead of continuous, so there should be no instanton proliferation problem. 

We emphasize that our results in two dimensions are limited to zero energy density (akin to Ref.~\cite{kim2016localization} but in one lower dimension). Formally, this is because exact emergence in two dimensions is limited to zero temperature~\cite{pace2023exact}. Intuitively, it is because there are finite-energy pointlike excitations consisting of endpoints of colored strings, and the propagation of these excitations across the system can change the Krylov sector label, on a timescale that is exponentially long in inverse perturbation strength~\cite{stahl2023topologically} but still finite. Stability at nonzero temperature requires going to three dimensions.

\emph{The model in three dimensions.}---%
In three dimensions, we construct a model of fluctuating membranes, whose boundaries themselves form closed loops. These boundaries have nonzero line tension, producing a diverging energy barrier between different RK states. 

The essential feature of the honeycomb lattice in the above construction is that three edges meet at every vertex. The analogue in three dimensions is to have three faces meet at every edge. One 3d lattice that satisfies this requirement is the rhombic dodecahedral honeycomb, but any 3d lattice with three faces meeting at every edge works. We place degrees of freedom on each face, with the same local Hilbert space as before. Define 
\begin{align}
A_e^{(\alpha)} = \prod_{f \in \pardag e} Z^{(\alpha)}_f,
\end{align}
which acts as $+1$ if edge $e$ has an even number of incident faces in state $\alpha$. Furthermore, define the kinetic and potential operators
\begin{align}
B_c^{K,\alpha} &= \prod_{f \in \partial c} X^{(\alpha)}_f, & B_c^{P,\alpha} &= \prod_{ f \in \partial c} (\ket{0} \bra{0} + \ket{\alpha} \bra{\alpha})_f, \nonumber
\end{align}
so that $B_c^{K,\alpha}$ flips all of the faces around cell $c$ when possible and $B_c^{P,\alpha}$ projects onto states on which $B_c^{K,\alpha}$ acts nontrivially. 

As before, we can define a classical Hamiltonian that enforces that $A_e^{(\alpha)} = 1$ for $\alpha>0$ for all edges. The emergent degrees of freedom become fluctuating colored membranes, and the Krylov labels describe the pattern of noncontractible membranes, similarly to Refs.~\cite{stahl2023topologically,khudorozhkov2024robust}. The kinetic operator $B_c^{K,\alpha}$ fluctuates an $\alpha$-colored membrane across cell $c$, and the potential operator $B_c^{P,\alpha}$ asks whether that process is allowed. 

The quantum Hamiltonian at its RK point,
\begin{equation}
H_0 = J \sum_e \sum_{\alpha = 1}^m \frac{1 \! - \! A_e^{(\alpha)}}{2} + K \sum_c \sum_{\alpha = 1}^m \frac{B_c^{P,\alpha} \! - \! B_c^{K,\alpha}}{2}
\end{equation}
has $(m-1)^L$ ground states, like the 2d model, each labeled by an RK label $\lambda$. In a slight abuse of notation, we will use $\ket{\Psi_\lambda}$ to also refer to this 3d state. For $\epsilon<0$, the true ground state is $\ket{\Psi_\varnothing}$ and the short-label states (defined similarly to above) have vanishing energy density. We refer to this model as the rainbow membrane soup, and analogously to the rainbow loop soup, the different short-label RK states should be steady states under generic perturbations up to a timescale that diverges in the thermodynamic limit. 

The major difference from the rainbow loop soup is that 
the $\ZZ_2^m$ symmetry is now a 2-form symmetry, and 2-form symmetries remain exact emergent even at nonzero temperature~\cite{pace2023exact}.
Meanwhile, changing Krylov sector using $k$-local moves requires going through a diverging number of intermediate states containing incomplete membranes, and membrane boundaries violate the 2-form symmetry (which, we recall, should be exact). Thus, we claim that every short-label RK state is stable even when connected to a nonzero-temperature bath.

The claim above, which follows from formal results on exact emergence~\cite{pace2023exact}, can be understood intuitively using a false-vacuum-decay argument. In order to change the Krylov sector, it is necessary to remove or insert an entire noncontractible membrane. When this deletion or insertion is incomplete, an extended loop of violated $A_e^{(\alpha)}$ operators surrounds the partial membrane. This means there is a linear energy barrier between the RK states. Meanwhile, the short-label RK states have vanishing energy density, so there is no `energy gain proportional to area' to compensate this  energy cost. 
If we go to nonzero temperature, there is a nonzero density of finite-energy excitations. However, the finite-energy excitations are local membrane boundaries with finite size, and the motion of local membrane boundaries cannot change the Krylov sector label. Changing Krylov sector label requires membrane boundaries of {\it diverging} size (and energy), and these should not appear at low temperatures. Accordingly, the Hilbert space shattering should remain stable at nonzero temperatures or in the presence of a nonzero density of local constraint violations. 
We emphasize that because we are relying on diverging energy barriers and not just vanishing matrix elements, the rainbow membrane soup should even be stable even when connected to a low-temperature bath, in contrast to MBL~\cite{mblbath}.

\emph{Discussion.}---%
We have proposed a new route to ergodicity breaking at infinite times in the thermodynamic limit in physically reasonable dimensions. In two dimensions, we have argued that it yields exponentially many infinite-time steady states at zero energy density, whereas in three dimensions similar arguments say that the exponentially numerous infinite-time steady states survive coupling to a nonzero temperature heat bath -- the first such example of which we are aware. We emphasize that our arguments are heuristic and do not constitute a proof. A proof would require, at a minimum, rigorously establishing the zero energy density and locally gapped nature of short-label false vacua, and demonstrating that the exact emergent symmetry is smoothly connected to the one in Eq.~\eqref{eqn:exactsymm}. In addition, while we have argued that our models evade the common instabilities that restore ergodicity on exponentially long timescales in previously studied models of Hilbert space shattering, it remains possible that there is some other, hitherto unappreciated instability to which our models are not immune. A rigorous proof of infinite-time stability (or lack thereof) is an obvious and important problem for future work. 

Another important question is how to extract Krylov sectors from the steady states. It is easy to extract labels from any states obeying the constraints of Eq.~\eqref{eqn:constraint}, but the true steady states will also include constraint-violating states, so some decoding may be necessary. This may take the form of some matching problem~\cite{dennis2002topological} for the 2d model, and simulated cooling~\cite{castelnovoChamon} 
for the 3d model. Whether these decoders would require further refinement remains to be seen.

Our models have a close connection to dimer models, quantum link models, and Rokhsar-Kivelson models, all of which have been extensively studied in the context of quantum spin liquids. While we have focused on the {\it dynamical} properties of our models, the {\it equilibrium} or {\it entanglement} properties of such models may be of independent interest, and would also constitute a worthy problem for future exploration. As mentioned, Ref.~\cite{zhang2024bicolor} has already shown interesting entanglement properties in the $m=2$ model, and models with $m>2$ might be even stranger, with $\exp(L)$ ground states instead of $\text{poly}(L)$, see also~\cite{balasubramanian2023entanglement}.

Finally, there is the question of how to realize the physics we have discussed in experiments, and indeed how to harness it, e.g., for the design of robust quantum memories. Experimental realization in synthetic systems seems the most direct route to laboratory realization, but given the close connection of our models to model Hamiltonians for quantum spin liquids, a quantum material realization may also be possible. This, too, is a worthy problem for future work. 

{\it Acknowledgements}: We thank Ethan Lake and Alexey Khudorozhkov for detailed feedback on the manuscript. This work was supported by the U.S. Department of Energy, Office of Science, Basic Energy Sciences, under Award DE-SC0021346.

\bibliography{robuster}

\begin{thebibliography}{37}%
\makeatletter
\providecommand \@ifxundefined [1]{%
 \@ifx{#1\undefined}
}%
\providecommand \@ifnum [1]{%
 \ifnum #1\expandafter \@firstoftwo
 \else \expandafter \@secondoftwo
 \fi
}%
\providecommand \@ifx [1]{%
 \ifx #1\expandafter \@firstoftwo
 \else \expandafter \@secondoftwo
 \fi
}%
\providecommand \natexlab [1]{#1}%
\providecommand \enquote  [1]{``#1''}%
\providecommand \bibnamefont  [1]{#1}%
\providecommand \bibfnamefont [1]{#1}%
\providecommand \citenamefont [1]{#1}%
\providecommand \href@noop [0]{\@secondoftwo}%
\providecommand \href [0]{\begingroup \@sanitize@url \@href}%
\providecommand \@href[1]{\@@startlink{#1}\@@href}%
\providecommand \@@href[1]{\endgroup#1\@@endlink}%
\providecommand \@sanitize@url [0]{\catcode `\\12\catcode `\$12\catcode
  `\&12\catcode `\#12\catcode `\^12\catcode `\_12\catcode `\%12\relax}%
\providecommand \@@startlink[1]{}%
\providecommand \@@endlink[0]{}%
\providecommand \url  [0]{\begingroup\@sanitize@url \@url }%
\providecommand \@url [1]{\endgroup\@href {#1}{\urlprefix }}%
\providecommand \urlprefix  [0]{URL }%
\providecommand \Eprint [0]{\href }%
\providecommand \doibase [0]{https://doi.org/}%
\providecommand \selectlanguage [0]{\@gobble}%
\providecommand \bibinfo  [0]{\@secondoftwo}%
\providecommand \bibfield  [0]{\@secondoftwo}%
\providecommand \translation [1]{[#1]}%
\providecommand \BibitemOpen [0]{}%
\providecommand \bibitemStop [0]{}%
\providecommand \bibitemNoStop [0]{.\EOS\space}%
\providecommand \EOS [0]{\spacefactor3000\relax}%
\providecommand \BibitemShut  [1]{\csname bibitem#1\endcsname}%
\let\auto@bib@innerbib\@empty
\bibitem [{\citenamefont {Nussinov}\ and\ \citenamefont
  {Ortiz}(2009)}]{Nussinov2007}%
  \BibitemOpen
  \bibfield  {author} {\bibinfo {author} {\bibfnamefont {Z.}~\bibnamefont
  {Nussinov}}\ and\ \bibinfo {author} {\bibfnamefont {G.}~\bibnamefont
  {Ortiz}},\ }\bibfield  {title} {\bibinfo {title} {A symmetry principle for
  topological quantum order},\ }\href
  {https://doi.org/doi.org/10.1016/j.aop.2008.11.002} {\bibfield  {journal}
  {\bibinfo  {journal} {Annals of Physics}\ }\textbf {\bibinfo {volume}
  {324}},\ \bibinfo {pages} {977} (\bibinfo {year} {2009})}\BibitemShut
  {NoStop}%
\bibitem [{\citenamefont {Gaiotto}\ \emph {et~al.}(2015)\citenamefont
  {Gaiotto}, \citenamefont {Kapustin}, \citenamefont {Seiberg},\ and\
  \citenamefont {Willett}}]{Gaiotto2015}%
  \BibitemOpen
  \bibfield  {author} {\bibinfo {author} {\bibfnamefont {D.}~\bibnamefont
  {Gaiotto}}, \bibinfo {author} {\bibfnamefont {A.}~\bibnamefont {Kapustin}},
  \bibinfo {author} {\bibfnamefont {N.}~\bibnamefont {Seiberg}},\ and\ \bibinfo
  {author} {\bibfnamefont {B.}~\bibnamefont {Willett}},\ }\bibfield  {title}
  {\bibinfo {title} {Generalized global symmetries},\ }\href
  {https://doi.org/10.1007/JHEP02(2015)172} {\bibfield  {journal} {\bibinfo
  {journal} {J. High Energ. Phys.}\ }\textbf {\bibinfo {volume} {2015}}\bibinfo
   {number} { (172)}}\BibitemShut {NoStop}%
\bibitem [{\citenamefont {McGreevy}(2023)}]{mcGreevy2023generalized}%
  \BibitemOpen
\bibfield  {number} {  }\bibfield  {author} {\bibinfo {author} {\bibfnamefont
  {J.}~\bibnamefont {McGreevy}},\ }\bibfield  {title} {\bibinfo {title}
  {Generalized symmetries in condensed matter},\ }\href
  {https://doi.org/10.1146/annurev-conmatphys-040721-021029} {\bibfield
  {journal} {\bibinfo  {journal} {Annual Review of Condensed Matter Physics}\
  }\textbf {\bibinfo {volume} {14}},\ \bibinfo {pages} {57–82} (\bibinfo
  {year} {2023})}\BibitemShut {NoStop}%
\bibitem [{\citenamefont {Yin}\ \emph {et~al.}(2024{\natexlab{a}})\citenamefont
  {Yin}, \citenamefont {Nandkishore},\ and\ \citenamefont {Lucas}}]{YNL}%
  \BibitemOpen
  \bibfield  {author} {\bibinfo {author} {\bibfnamefont {C.}~\bibnamefont
  {Yin}}, \bibinfo {author} {\bibfnamefont {R.}~\bibnamefont {Nandkishore}},\
  and\ \bibinfo {author} {\bibfnamefont {A.}~\bibnamefont {Lucas}},\
  }\href@noop {} {\bibinfo {title} {Eigenstate localization in a many-body
  quantum system}} (\bibinfo {year} {2024}{\natexlab{a}}),\ \Eprint
  {https://arxiv.org/abs/2405.12279} {arXiv:2405.12279 [cond-mat.stat-mech]}
  \BibitemShut {NoStop}%
\bibitem [{\citenamefont {Baxter}(2016)}]{Baxter}%
  \BibitemOpen
  \bibfield  {author} {\bibinfo {author} {\bibfnamefont {R.~J.}\ \bibnamefont
  {Baxter}},\ }\href@noop {} {\emph {\bibinfo {title} {Exactly solved models in
  statistical mechanics}}}\ (\bibinfo  {publisher} {Elsevier},\ \bibinfo {year}
  {2016})\BibitemShut {NoStop}%
\bibitem [{\citenamefont {Nandkishore}\ and\ \citenamefont
  {Huse}(2015)}]{MBLARCMP}%
  \BibitemOpen
  \bibfield  {author} {\bibinfo {author} {\bibfnamefont {R.}~\bibnamefont
  {Nandkishore}}\ and\ \bibinfo {author} {\bibfnamefont {D.~A.}\ \bibnamefont
  {Huse}},\ }\bibfield  {title} {\bibinfo {title} {Many-body localization and
  thermalization in quantum statistical mechanics},\ }\href
  {https://doi.org/10.1146/annurev-conmatphys-031214-014726} {\bibfield
  {journal} {\bibinfo  {journal} {Annu. Rev. Condens. Matter Phys.}\ }\textbf
  {\bibinfo {volume} {6}},\ \bibinfo {pages} {15} (\bibinfo {year}
  {2015})}\BibitemShut {NoStop}%
\bibitem [{\citenamefont {Abanin}\ \emph {et~al.}(2019)\citenamefont {Abanin},
  \citenamefont {Altman}, \citenamefont {Bloch},\ and\ \citenamefont
  {Serbyn}}]{MBLRMP}%
  \BibitemOpen
  \bibfield  {author} {\bibinfo {author} {\bibfnamefont {D.~A.}\ \bibnamefont
  {Abanin}}, \bibinfo {author} {\bibfnamefont {E.}~\bibnamefont {Altman}},
  \bibinfo {author} {\bibfnamefont {I.}~\bibnamefont {Bloch}},\ and\ \bibinfo
  {author} {\bibfnamefont {M.}~\bibnamefont {Serbyn}},\ }\bibfield  {title}
  {\bibinfo {title} {Colloquium: Many-body localization, thermalization, and
  entanglement},\ }\href {https://doi.org/10.1103/RevModPhys.91.021001}
  {\bibfield  {journal} {\bibinfo  {journal} {Rev. Mod. Phys.}\ }\textbf
  {\bibinfo {volume} {91}},\ \bibinfo {pages} {021001} (\bibinfo {year}
  {2019})}\BibitemShut {NoStop}%
\bibitem [{\citenamefont {Nandkishore}\ and\ \citenamefont
  {Gopalakrishnan}(2017)}]{mblbath}%
  \BibitemOpen
  \bibfield  {author} {\bibinfo {author} {\bibfnamefont {R.}~\bibnamefont
  {Nandkishore}}\ and\ \bibinfo {author} {\bibfnamefont {S.}~\bibnamefont
  {Gopalakrishnan}},\ }\bibfield  {title} {\bibinfo {title} {Many body
  localized systems weakly coupled to baths},\ }\href
  {https://doi.org/10.1002/andp.201600181} {\bibfield  {journal} {\bibinfo
  {journal} {Annalen der Physik}\ }\textbf {\bibinfo {volume} {529}},\ \bibinfo
  {pages} {1600181} (\bibinfo {year} {2017})}\BibitemShut {NoStop}%
\bibitem [{\citenamefont {Imbrie}(2016)}]{Imbrie}%
  \BibitemOpen
  \bibfield  {author} {\bibinfo {author} {\bibfnamefont {J.~Z.}\ \bibnamefont
  {Imbrie}},\ }\bibfield  {title} {\bibinfo {title} {On many-body localization
  for quantum spin chains},\ }\href {https://doi.org/10.1007/s10955-016-1508-x}
  {\bibfield  {journal} {\bibinfo  {journal} {Journal of Statistical Physics}\
  }\textbf {\bibinfo {volume} {163}},\ \bibinfo {pages} {998} (\bibinfo {year}
  {2016})}\BibitemShut {NoStop}%
\bibitem [{\citenamefont {Khemani}\ \emph {et~al.}(2020)\citenamefont
  {Khemani}, \citenamefont {Hermele},\ and\ \citenamefont
  {Nandkishore}}]{khemani2020shattering}%
  \BibitemOpen
  \bibfield  {author} {\bibinfo {author} {\bibfnamefont {V.}~\bibnamefont
  {Khemani}}, \bibinfo {author} {\bibfnamefont {M.}~\bibnamefont {Hermele}},\
  and\ \bibinfo {author} {\bibfnamefont {R.}~\bibnamefont {Nandkishore}},\
  }\bibfield  {title} {\bibinfo {title} {Localization from {H}ilbert space
  shattering: From theory to physical realizations},\ }\href
  {https://doi.org/10.1103/PhysRevB.101.174204} {\bibfield  {journal} {\bibinfo
   {journal} {Physical Review B}\ }\textbf {\bibinfo {volume} {101}},\ \bibinfo
  {pages} {174204} (\bibinfo {year} {2020})}\BibitemShut {NoStop}%
\bibitem [{\citenamefont {Sala}\ \emph {et~al.}(2020)\citenamefont {Sala},
  \citenamefont {Rakovszky}, \citenamefont {Verresen}, \citenamefont {Knap},\
  and\ \citenamefont {Pollmann}}]{Sala2020}%
  \BibitemOpen
  \bibfield  {author} {\bibinfo {author} {\bibfnamefont {P.}~\bibnamefont
  {Sala}}, \bibinfo {author} {\bibfnamefont {T.}~\bibnamefont {Rakovszky}},
  \bibinfo {author} {\bibfnamefont {R.}~\bibnamefont {Verresen}}, \bibinfo
  {author} {\bibfnamefont {M.}~\bibnamefont {Knap}},\ and\ \bibinfo {author}
  {\bibfnamefont {F.}~\bibnamefont {Pollmann}},\ }\bibfield  {title} {\bibinfo
  {title} {Ergodicity breaking arising from {H}ilbert space fragmentation in
  dipole-conserving {H}amiltonians},\ }\href
  {https://doi.org/10.1103/PhysRevX.10.011047} {\bibfield  {journal} {\bibinfo
  {journal} {Physical Review X}\ }\textbf {\bibinfo {volume} {10}},\ \bibinfo
  {pages} {011047} (\bibinfo {year} {2020})}\BibitemShut {NoStop}%
\bibitem [{\citenamefont {Stephen}\ \emph {et~al.}(2024)\citenamefont
  {Stephen}, \citenamefont {Hart},\ and\ \citenamefont
  {Nandkishore}}]{stephen2022ergodicity}%
  \BibitemOpen
  \bibfield  {author} {\bibinfo {author} {\bibfnamefont {D.~T.}\ \bibnamefont
  {Stephen}}, \bibinfo {author} {\bibfnamefont {O.}~\bibnamefont {Hart}},\ and\
  \bibinfo {author} {\bibfnamefont {R.~M.}\ \bibnamefont {Nandkishore}},\
  }\bibfield  {title} {\bibinfo {title} {Ergodicity breaking provably robust to
  arbitrary perturbations},\ }\href
  {https://doi.org/10.1103/PhysRevLett.132.040401} {\bibfield  {journal}
  {\bibinfo  {journal} {Phys. Rev. Lett.}\ }\textbf {\bibinfo {volume} {132}},\
  \bibinfo {pages} {040401} (\bibinfo {year} {2024})}\BibitemShut {NoStop}%
\bibitem [{\citenamefont {Stahl}\ \emph {et~al.}(2024)\citenamefont {Stahl},
  \citenamefont {Nandkishore},\ and\ \citenamefont
  {Hart}}]{stahl2023topologically}%
  \BibitemOpen
  \bibfield  {author} {\bibinfo {author} {\bibfnamefont {C.}~\bibnamefont
  {Stahl}}, \bibinfo {author} {\bibfnamefont {R.}~\bibnamefont {Nandkishore}},\
  and\ \bibinfo {author} {\bibfnamefont {O.}~\bibnamefont {Hart}},\ }\bibfield
  {title} {\bibinfo {title} {{Topologically stable ergodicity breaking from
  emergent higher-form symmetries in generalized quantum loop models}},\ }\href
  {https://doi.org/10.21468/SciPostPhys.16.3.068} {\bibfield  {journal}
  {\bibinfo  {journal} {SciPost Phys.}\ }\textbf {\bibinfo {volume} {16}},\
  \bibinfo {pages} {068} (\bibinfo {year} {2024})}\BibitemShut {NoStop}%
\bibitem [{\citenamefont {Balasubramanian}\ \emph {et~al.}(2024)\citenamefont
  {Balasubramanian}, \citenamefont {Gopalakrishnan}, \citenamefont
  {Khudorozhkov},\ and\ \citenamefont {Lake}}]{Balasubramanian2024glassy}%
  \BibitemOpen
  \bibfield  {author} {\bibinfo {author} {\bibfnamefont {S.}~\bibnamefont
  {Balasubramanian}}, \bibinfo {author} {\bibfnamefont {S.}~\bibnamefont
  {Gopalakrishnan}}, \bibinfo {author} {\bibfnamefont {A.}~\bibnamefont
  {Khudorozhkov}},\ and\ \bibinfo {author} {\bibfnamefont {E.}~\bibnamefont
  {Lake}},\ }\bibfield  {title} {\bibinfo {title} {Glassy word problems:
  Ultraslow relaxation, {H}ilbert space jamming, and computational
  complexity},\ }\href {https://doi.org/10.1103/PhysRevX.14.021034} {\bibfield
  {journal} {\bibinfo  {journal} {Phys. Rev. X}\ }\textbf {\bibinfo {volume}
  {14}},\ \bibinfo {pages} {021034} (\bibinfo {year} {2024})}\BibitemShut
  {NoStop}%
\bibitem [{\citenamefont {Khudorozhkov}\ \emph {et~al.}(2024)\citenamefont
  {Khudorozhkov}, \citenamefont {Stahl}, \citenamefont {Hart},\ and\
  \citenamefont {Nandkishore}}]{khudorozhkov2024robust}%
  \BibitemOpen
  \bibfield  {author} {\bibinfo {author} {\bibfnamefont {A.}~\bibnamefont
  {Khudorozhkov}}, \bibinfo {author} {\bibfnamefont {C.}~\bibnamefont {Stahl}},
  \bibinfo {author} {\bibfnamefont {O.}~\bibnamefont {Hart}},\ and\ \bibinfo
  {author} {\bibfnamefont {R.}~\bibnamefont {Nandkishore}},\ }\href@noop {}
  {\bibinfo {title} {Robust {H}ilbert space fragmentation in group-valued loop
  models}} (\bibinfo {year} {2024}),\ \Eprint
  {https://arxiv.org/abs/arXiv:2406.19386} {arXiv:2406.19386} \BibitemShut
  {NoStop}%
\bibitem [{\citenamefont {Nandkishore}\ and\ \citenamefont
  {Hermele}(2019)}]{fractonarcmp}%
  \BibitemOpen
  \bibfield  {author} {\bibinfo {author} {\bibfnamefont {R.~M.}\ \bibnamefont
  {Nandkishore}}\ and\ \bibinfo {author} {\bibfnamefont {M.}~\bibnamefont
  {Hermele}},\ }\bibfield  {title} {\bibinfo {title} {Fractons},\ }\href
  {https://doi.org/10.1146/annurev-conmatphys-031218-013604} {\bibfield
  {journal} {\bibinfo  {journal} {Annual Review of Condensed Matter Physics}\
  }\textbf {\bibinfo {volume} {10}},\ \bibinfo {pages} {295} (\bibinfo {year}
  {2019})}\BibitemShut {NoStop}%
\bibitem [{\citenamefont {Gromov}\ and\ \citenamefont
  {Radzihovsky}(2024)}]{fractonrmp}%
  \BibitemOpen
  \bibfield  {author} {\bibinfo {author} {\bibfnamefont {A.}~\bibnamefont
  {Gromov}}\ and\ \bibinfo {author} {\bibfnamefont {L.}~\bibnamefont
  {Radzihovsky}},\ }\bibfield  {title} {\bibinfo {title} {Colloquium: Fracton
  matter},\ }\href {https://doi.org/10.1103/RevModPhys.96.011001} {\bibfield
  {journal} {\bibinfo  {journal} {Rev. Mod. Phys.}\ }\textbf {\bibinfo {volume}
  {96}},\ \bibinfo {pages} {011001} (\bibinfo {year} {2024})}\BibitemShut
  {NoStop}%
\bibitem [{\citenamefont {Kim}\ and\ \citenamefont
  {Haah}(2016)}]{kim2016localization}%
  \BibitemOpen
  \bibfield  {author} {\bibinfo {author} {\bibfnamefont {I.~H.}\ \bibnamefont
  {Kim}}\ and\ \bibinfo {author} {\bibfnamefont {J.}~\bibnamefont {Haah}},\
  }\bibfield  {title} {\bibinfo {title} {Localization from superselection rules
  in translationally invariant systems},\ }\href
  {https://doi.org/10.1103/physrevlett.116.027202} {\bibfield  {journal}
  {\bibinfo  {journal} {Physical Review Letters}\ }\textbf {\bibinfo {volume}
  {116}},\ \bibinfo {pages} {027202} (\bibinfo {year} {2016})}\BibitemShut
  {NoStop}%
\bibitem [{\citenamefont {Prem}\ \emph {et~al.}(2017)\citenamefont {Prem},
  \citenamefont {Haah},\ and\ \citenamefont {Nandkishore}}]{prem2017glassy}%
  \BibitemOpen
  \bibfield  {author} {\bibinfo {author} {\bibfnamefont {A.}~\bibnamefont
  {Prem}}, \bibinfo {author} {\bibfnamefont {J.}~\bibnamefont {Haah}},\ and\
  \bibinfo {author} {\bibfnamefont {R.}~\bibnamefont {Nandkishore}},\
  }\bibfield  {title} {\bibinfo {title} {Glassy quantum dynamics in translation
  invariant fracton models},\ }\href
  {https://doi.org/10.1103/PhysRevB.95.155133} {\bibfield  {journal} {\bibinfo
  {journal} {Phys. Rev. B}\ }\textbf {\bibinfo {volume} {95}},\ \bibinfo
  {pages} {155133} (\bibinfo {year} {2017})}\BibitemShut {NoStop}%
\bibitem [{\citenamefont {Siva}\ and\ \citenamefont
  {Yoshida}(2017)}]{siva2017marginally}%
  \BibitemOpen
  \bibfield  {author} {\bibinfo {author} {\bibfnamefont {K.}~\bibnamefont
  {Siva}}\ and\ \bibinfo {author} {\bibfnamefont {B.}~\bibnamefont {Yoshida}},\
  }\bibfield  {title} {\bibinfo {title} {Topological order and memory time in
  marginally-self-correcting quantum memory},\ }\href
  {https://doi.org/10.1103/PhysRevA.95.032324} {\bibfield  {journal} {\bibinfo
  {journal} {Phys. Rev. A}\ }\textbf {\bibinfo {volume} {95}},\ \bibinfo
  {pages} {032324} (\bibinfo {year} {2017})}\BibitemShut {NoStop}%
\bibitem [{\citenamefont {Pace}\ and\ \citenamefont
  {Wen}(2023)}]{pace2023exact}%
  \BibitemOpen
  \bibfield  {author} {\bibinfo {author} {\bibfnamefont {S.~D.}\ \bibnamefont
  {Pace}}\ and\ \bibinfo {author} {\bibfnamefont {X.-G.}\ \bibnamefont {Wen}},\
  }\bibfield  {title} {\bibinfo {title} {Exact emergent higher-form symmetries
  in bosonic lattice models},\ }\href
  {https://doi.org/10.1103/physrevb.108.195147} {\bibfield  {journal} {\bibinfo
   {journal} {Physical Review B}\ }\textbf {\bibinfo {volume} {108}},\ \bibinfo
  {pages} {195147} (\bibinfo {year} {2023})}\BibitemShut {NoStop}%
\bibitem [{\citenamefont {Caha}\ and\ \citenamefont
  {Nagaj}(2018)}]{caha2018pairflip}%
  \BibitemOpen
  \bibfield  {author} {\bibinfo {author} {\bibfnamefont {L.}~\bibnamefont
  {Caha}}\ and\ \bibinfo {author} {\bibfnamefont {D.}~\bibnamefont {Nagaj}},\
  }\href@noop {} {\bibinfo {title} {The pair-flip model: a very entangled
  translationally invariant spin chain}} (\bibinfo {year} {2018}),\ \Eprint
  {https://arxiv.org/abs/arXiv:1805.07168} {arXiv:1805.07168} \BibitemShut
  {NoStop}%
\bibitem [{\citenamefont {Moudgalya}\ and\ \citenamefont
  {Motrunich}(2022)}]{moudgalya2022hilbert}%
  \BibitemOpen
  \bibfield  {author} {\bibinfo {author} {\bibfnamefont {S.}~\bibnamefont
  {Moudgalya}}\ and\ \bibinfo {author} {\bibfnamefont {O.~I.}\ \bibnamefont
  {Motrunich}},\ }\bibfield  {title} {\bibinfo {title} {Hilbert space
  fragmentation and commutant algebras},\ }\href
  {https://doi.org/10.1103/PhysRevX.12.011050} {\bibfield  {journal} {\bibinfo
  {journal} {Physical Review X}\ }\textbf {\bibinfo {volume} {12}},\ \bibinfo
  {pages} {011050} (\bibinfo {year} {2022})}\BibitemShut {NoStop}%
\bibitem [{Note1()}]{Note1}%
  \BibitemOpen
  \bibinfo {note} {In order to obtain labels in one-to-one correspondence with
  Krylov sectors, identify labels that are equivalent up to periodic
  translations, unless the word has length $L$. See Ref.~\cite
  {khudorozhkov2024robust} for further details.}\BibitemShut {Stop}%
\bibitem [{\citenamefont {Rokhsar}\ and\ \citenamefont {Kivelson}(1988)}]{RK}%
  \BibitemOpen
  \bibfield  {author} {\bibinfo {author} {\bibfnamefont {D.~S.}\ \bibnamefont
  {Rokhsar}}\ and\ \bibinfo {author} {\bibfnamefont {S.~A.}\ \bibnamefont
  {Kivelson}},\ }\bibfield  {title} {\bibinfo {title} {Superconductivity and
  the quantum hard-core dimer gas},\ }\href
  {https://doi.org/10.1103/PhysRevLett.61.2376} {\bibfield  {journal} {\bibinfo
   {journal} {Phys. Rev. Lett.}\ }\textbf {\bibinfo {volume} {61}},\ \bibinfo
  {pages} {2376} (\bibinfo {year} {1988})}\BibitemShut {NoStop}%
\bibitem [{\citenamefont {Henley}(2004)}]{henley2004RK}%
  \BibitemOpen
  \bibfield  {author} {\bibinfo {author} {\bibfnamefont {C.~L.}\ \bibnamefont
  {Henley}},\ }\bibfield  {title} {\bibinfo {title} {From classical to quantum
  dynamics at {R}okhsar–{K}ivelson points},\ }\href
  {https://doi.org/10.1088/0953-8984/16/11/045} {\bibfield  {journal} {\bibinfo
   {journal} {Journal of Physics: Condensed Matter}\ }\textbf {\bibinfo
  {volume} {16}},\ \bibinfo {pages} {S891–S898} (\bibinfo {year}
  {2004})}\BibitemShut {NoStop}%
\bibitem [{\citenamefont {Zhang}(2024)}]{zhang2024bicolor}%
  \BibitemOpen
  \bibfield  {author} {\bibinfo {author} {\bibfnamefont {Z.}~\bibnamefont
  {Zhang}},\ }\bibfield  {title} {\bibinfo {title} {Bicolor loop models and
  their long range entanglement},\ }\href
  {https://doi.org/10.22331/q-2024-02-29-1268} {\bibfield  {journal} {\bibinfo
  {journal} {Quantum}\ }\textbf {\bibinfo {volume} {8}},\ \bibinfo {pages}
  {1268} (\bibinfo {year} {2024})}\BibitemShut {NoStop}%
\bibitem [{\citenamefont {Kitaev}(2003)}]{kitaev2003toric}%
  \BibitemOpen
  \bibfield  {author} {\bibinfo {author} {\bibfnamefont {A.~{\relax Yu}.}\
  \bibnamefont {Kitaev}},\ }\bibfield  {title} {\bibinfo {title}
  {Fault-tolerant quantum computation by anyons},\ }\href
  {https://doi.org/https://doi.org/10.1016/S0003-4916(02)00018-0} {\bibfield
  {journal} {\bibinfo  {journal} {Annals of Physics}\ }\textbf {\bibinfo
  {volume} {303}},\ \bibinfo {pages} {2} (\bibinfo {year} {2003})}\BibitemShut
  {NoStop}%
\bibitem [{Note2()}]{Note2}%
  \BibitemOpen
  \bibinfo {note} {Strictly the RK states are the steady states of the RK
  Hamiltonian $H_0$, not the perturbed Hamiltonian $H$. The true false vacua
  will be connected to these RK states by a Schrieffer-Wolff transformation
  \cite {SW}.}\BibitemShut {Stop}%
\bibitem [{sup()}]{supplement}%
  \BibitemOpen
  \href@noop {} {}\bibinfo {note} {See Supplemental Material (appended), for
  details of the numerical simulations.}\BibitemShut {Stop}%
\bibitem [{Note3()}]{Note3}%
  \BibitemOpen
  \bibinfo {note} {However, this is by no means proven, and placing this
  (highly plausible) result on firm footing is a key step towards a
  rigorization of our results}\BibitemShut {NoStop}%
\bibitem [{\citenamefont {Yin}\ \emph {et~al.}(2024{\natexlab{b}})\citenamefont
  {Yin}, \citenamefont {Surace},\ and\ \citenamefont {Lucas}}]{YSL}%
  \BibitemOpen
  \bibfield  {author} {\bibinfo {author} {\bibfnamefont {C.}~\bibnamefont
  {Yin}}, \bibinfo {author} {\bibfnamefont {F.~M.}\ \bibnamefont {Surace}},\
  and\ \bibinfo {author} {\bibfnamefont {A.}~\bibnamefont {Lucas}},\
  }\href@noop {} {\bibinfo {title} {Theory of metastable states in many-body
  quantum systems}} (\bibinfo {year} {2024}{\natexlab{b}}),\ \Eprint
  {https://arxiv.org/abs/2408.05261} {arXiv:2408.05261 [math-ph]} \BibitemShut
  {NoStop}%
\bibitem [{\citenamefont {Polyakov}(1987)}]{Polyakov}%
  \BibitemOpen
  \bibfield  {author} {\bibinfo {author} {\bibfnamefont {A.~M.}\ \bibnamefont
  {Polyakov}},\ }\href@noop {} {\emph {\bibinfo {title} {Gauge fields and
  strings}}}\ (\bibinfo  {publisher} {Taylor \& Francis},\ \bibinfo {year}
  {1987})\BibitemShut {NoStop}%
\bibitem [{\citenamefont {Dennis}\ \emph {et~al.}(2002)\citenamefont {Dennis},
  \citenamefont {Kitaev}, \citenamefont {Landahl},\ and\ \citenamefont
  {Preskill}}]{dennis2002topological}%
  \BibitemOpen
  \bibfield  {author} {\bibinfo {author} {\bibfnamefont {E.}~\bibnamefont
  {Dennis}}, \bibinfo {author} {\bibfnamefont {A.}~\bibnamefont {Kitaev}},
  \bibinfo {author} {\bibfnamefont {A.}~\bibnamefont {Landahl}},\ and\ \bibinfo
  {author} {\bibfnamefont {J.}~\bibnamefont {Preskill}},\ }\bibfield  {title}
  {\bibinfo {title} {Topological quantum memory},\ }\href
  {https://doi.org/10.1063/1.1499754} {\bibfield  {journal} {\bibinfo
  {journal} {Journal of Mathematical Physics}\ }\textbf {\bibinfo {volume}
  {43}},\ \bibinfo {pages} {4452–4505} (\bibinfo {year} {2002})}\BibitemShut
  {NoStop}%
\bibitem [{\citenamefont {Castelnovo}\ and\ \citenamefont
  {Chamon}(2008)}]{castelnovoChamon}%
  \BibitemOpen
  \bibfield  {author} {\bibinfo {author} {\bibfnamefont {C.}~\bibnamefont
  {Castelnovo}}\ and\ \bibinfo {author} {\bibfnamefont {C.}~\bibnamefont
  {Chamon}},\ }\bibfield  {title} {\bibinfo {title} {Topological order in a
  three-dimensional toric code at finite temperature},\ }\href
  {https://doi.org/10.1103/PhysRevB.78.155120} {\bibfield  {journal} {\bibinfo
  {journal} {Phys. Rev. B}\ }\textbf {\bibinfo {volume} {78}},\ \bibinfo
  {pages} {155120} (\bibinfo {year} {2008})}\BibitemShut {NoStop}%
\bibitem [{\citenamefont {Balasubramanian}\ \emph {et~al.}(2023)\citenamefont
  {Balasubramanian}, \citenamefont {Lake},\ and\ \citenamefont
  {Choi}}]{balasubramanian2023entanglement}%
  \BibitemOpen
  \bibfield  {author} {\bibinfo {author} {\bibfnamefont {S.}~\bibnamefont
  {Balasubramanian}}, \bibinfo {author} {\bibfnamefont {E.}~\bibnamefont
  {Lake}},\ and\ \bibinfo {author} {\bibfnamefont {S.}~\bibnamefont {Choi}},\
  }\href@noop {} {\bibinfo {title} {2d {H}amiltonians with exotic bipartite and
  topological entanglement}} (\bibinfo {year} {2023}),\ \Eprint
  {https://arxiv.org/abs/arXiv:2305.07028} {arXiv:2305.07028} \BibitemShut
  {NoStop}%
\bibitem [{\citenamefont {Bravyi}\ \emph {et~al.}(2011)\citenamefont {Bravyi},
  \citenamefont {DiVincenzo},\ and\ \citenamefont {Loss}}]{SW}%
  \BibitemOpen
  \bibfield  {author} {\bibinfo {author} {\bibfnamefont {S.}~\bibnamefont
  {Bravyi}}, \bibinfo {author} {\bibfnamefont {D.~P.}\ \bibnamefont
  {DiVincenzo}},\ and\ \bibinfo {author} {\bibfnamefont {D.}~\bibnamefont
  {Loss}},\ }\bibfield  {title} {\bibinfo {title} {Schrieffer--{W}olff
  transformation for quantum many-body systems},\ }\href
  {https://doi.org/10.1016/j.aop.2011.06.004} {\bibfield  {journal} {\bibinfo
  {journal} {Annals of Physics}\ }\textbf {\bibinfo {volume} {326}},\ \bibinfo
  {pages} {2793} (\bibinfo {year} {2011})}\BibitemShut {NoStop}%
\end{thebibliography}%

\cleardoublepage
\newpage

\onecolumngrid
\begin{center}
\textbf{\large Supplemental Material for ``Towards absolutely stable ergodicity breaking in finite dimensions''}
\vskip 0.4cm
{Charles{\;\,}Stahl,{\;\,}Oliver{\;\,}Hart,{\;\,}and{\;\,}Rahul{\;\,}M.{\;\,}Nandkishore}
\vskip 0.1cm
{\fontsize{9.5pt}{11.5pt}\selectfont\emph{Department of Physics and Center for Theory of Quantum Matter,\\[-0.75pt] University of Colorado, Boulder, Colorado 80309, USA}}\\[-0.75pt]
{\fontsize{9pt}{11pt}\selectfont(Dated: October 10, 2024)}
\vskip 0.75cm
\end{center}
\twocolumngrid

In the main text we claim that the short-label RK states are separated by zero energy density, even considering generic Hamiltonian perturbations. Here we supply some simple numeric arguments to back up this claim. First, we show that this is true away from the RK point but without perturbations, and then we show that this remains true for a particular non-RK perturbation. In both cases we work to first order in perturbation theory, so that the energy can be estimated using the expectation value of local operators in the classical statistical-mechanical ensemble corresponding to a particular Krylov sector~\cite{henley2004RK}, avoiding exact diagonalization. Note that the arguments are therefore illustrative rather than rigorous.

Let us start with the Hamiltonian
\begin{equation}
H_\RK[V=K+\epsilon, K]
\end{equation}
so that the RK states are no longer energy eigenstates. As described in the main text, we can treat 
\begin{equation}
\delta H = \epsilon \sum_f \sum_{\alpha=1}^m B_f^{P,\alpha}
\end{equation}
as a perturbation to determine the new energies. In the full treatment, we would also then go on to calculate the true steady states, which would be decorated versions of the RK states. Instead, for our purposes we will just calculate the energy shift to first order in perturbation theory, which is
\begin{align}
\delta E_\lambda &= \langle \Psi_\lambda | \delta H | \Psi_\lambda \rangle \nonumber\\
&= \epsilon L^2 m \langle B_f^{P,\alpha} \rangle_\lambda \label{eqn:perturb}
\end{align}
to the energy, where $\langle \ \cdot \ \rangle_\lambda$ denotes the expectation value in the state $\ket{\Psi_\lambda}$, averaged over $f$ and $\alpha$. Let us define $\rho^P_\lambda \equiv \langle B_f^{P,\alpha} \rangle_\lambda$ as the density of flippable faces.

It is simple to measure $\rho^P_\lambda$ by Monte-Carlo sampling from the classical ensemble consisting of every product state in $\mathcal{K}_\lambda$ with equal weight. 
To do this, we start in a known state in $\mathcal{K}_\lambda$, consisting of noncontractible loops forming the label lambda with no additional noncontractible loops. Then, we apply $B_f^{K,\alpha}$ operators (choosing $f$ and $\alpha$ randomly from the uniform distribution) until the state thermalizes with respect to $\langle B_f^{P,\alpha} \rangle$. Then, to resample from the ensemble, we continue to evolve the state past its autocorrelation time and measure $\langle B_f^{K,\alpha}\rangle$. We choose to work with $m=3$ for the numerics.

\begin{figure}
    \centering
    \includegraphics[width=\linewidth]{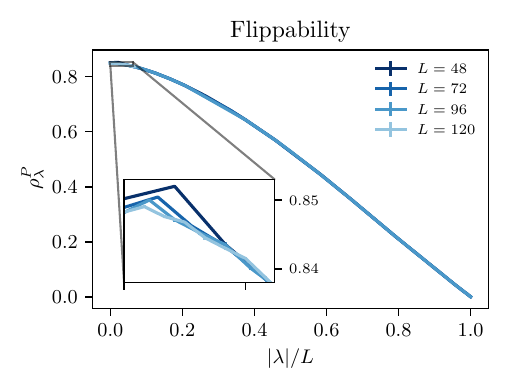}
    \caption{To first order in perturbation theory, the expectation value $\langle B_f^{P,\alpha} \rangle$ in the RK state $\Psi_\lambda$ is the fraction of flippable plaquettes in the classical ensemble of the Krylov sector $\mathcal{K}_\lambda$.}
    \label{fig:flippability}
\end{figure}

We expect that $\rho_\lambda^P$ should depend only on $|\lambda|$, so we choose representative $\lambda$ for different lengths; in particular, we choose a label that looks like $1212\cdots 12$, appending a $3$ to achieve odd $|\lambda|$. We have also verified that the results for random (nonrepeating) labels are indistinguishable. We report the results in Fig.~\ref{fig:flippability}. The curves are nearly identical when plotted as a function of $|\lambda| / L$, suggesting that $\rho_\lambda^P$ is  a function of that combination only, not $|\lambda|$ or $L$ individually. In particular, this suggests that $\rho_\lambda^P$ approaches a constant value for $|\lambda|/L \rightarrow 0$, meaning all short-label states have the same energy density. Note that Fig.~\ref{fig:flippability} exhibits nonmonotonic behavior for $|\lambda| = 0,1$. This appears to be a finite-size effect; as system size increases the flippability difference between $|\lambda|=0$ and 1 decreases.

We use a similar approach to explore the behavior of the RK states when subjected to generic perturbations. While there are many perturbations we could try, we choose a paramagnetic perturbation because, intuitively, states with many $\alpha$ characters in their label will also have many spins in the $\alpha$ state. The perturbation we consider is the paramagnetic field 
\begin{equation}
H_\text{para} = \sum_{\alpha=0}^m h_\alpha P^\alpha,
\end{equation}
where $P^\alpha \equiv \ket{\alpha} \bra{\alpha}$ is the projector onto state $\ket{\alpha}$. Then the local expectation value we are interested in is just $\langle P^\alpha \rangle_\alpha$. We now restrict to $\lambda$  of the form $1212\cdots 12$ with even $|\lambda|$. We expect that when $\lambda$ is short there are the same number of $1$, $2$, and $3$ spins, while as $|\lambda|$ increases the number of $1$ and $2$ spins increases and the numbers of $0$ and $3$ spins decreases.

\begin{figure}
    \centering
    \includegraphics[]{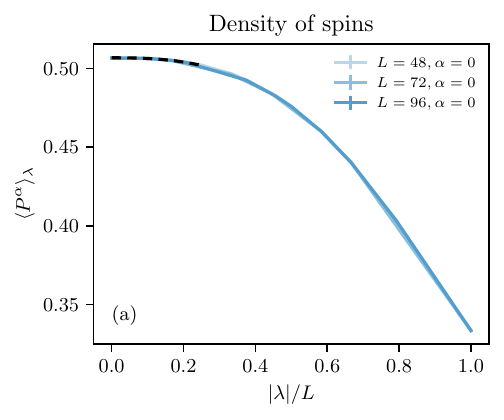}
    \includegraphics[]{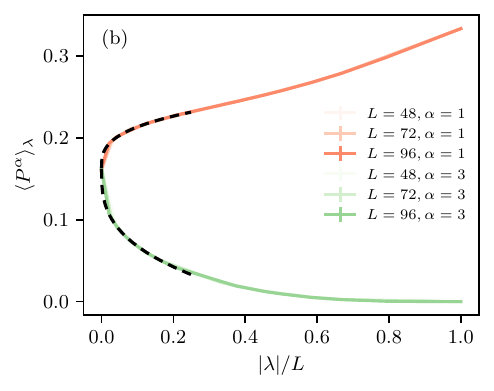}
    \caption{The expectation value $\langle P^\alpha \rangle_\lambda$, along with power-law best-fit curves described in the text, denoted by dashed lines.}
    \label{fig:Nspins}
\end{figure}

In Fig.~\ref{fig:Nspins}(a) we see that the number of $\alpha=0$ spins does in fact decrease with longer $\lambda$. In fact, the data are clean enough to fit to power law decays. Fitting to a power-law decay at short label lengths for $L=96$, 
\begin{equation}
\langle P^0 \rangle_\lambda = \rho_0 + b_1 \, \left(\frac{|\lambda|}{L}\right)^{\xi_1},
\end{equation}
we find
\begin{align}
\rho_0 &= 0.50672(3)\\
b_1 &= -0.201(3)\\
\xi_1 &= 2.70(2).
\end{align}
For label lengths approaching system size, it is reasonable for $\langle P^0 \rangle_\lambda$ to approach $1/3$ because, in a fully-packed configuration, one third of the spins are in state $|0\rangle$ (see Fig.~\ref{fig:symloops}).

In Fig.~\ref{fig:Nspins}(b) we see the analogous plots for $\alpha=1$ and $\alpha=3$. As a sanity check, the curves start with $\langle P^1 \rangle_\lambda = \langle P^3 \rangle_\lambda$ for $|\lambda|=0$ and end with $\langle P^3 \rangle_\lambda=0$. Fitting to the same power law at small label length,
for $\alpha=1$ we find 
\begin{align}
\rho_0 &= 0.1628(5)\\
b_1 &= 0.109(1)\\
\xi_1 &= 0.333(7),
\end{align}
while for $\alpha=3$ we find 
\begin{align}
\rho_0 &= 0.1618(5)\\
b_1 &= -0.207(1)\\
\xi_1 &= 0.343(4).
\end{align}

While a full understanding of the behavior of these curves would be interesting, the important conclusion is that the expectation values are all well-behaved as $|\lambda|/L \rightarrow 0$. This means that energy densities above the true ground state vanish in the thermodynamic limit for RK states with $|\lambda|/L \rightarrow 0$.

\end{document}